\documentclass[11pt,letterpaper,twoside]{article}
\usepackage[centertags]{amsmath}
\usepackage{graphicx,indentfirst,amsmath,amsfonts,amssymb,amsthm,newlfont}
\font\Goth=yinitas scaled \magstep0
\font\goth=ygoth scaled 1200
\newcommand{\Gth}[1]{\lower2mm\hbox{\Goth #1}}

\usepackage[latin1]{inputenc}
\def\al{\alpha}

\def\be{\beta}

\def\l1{{\lambda}_1}

\newcommand{\f}{\frac}

\def\x1{{\xi }_{xx}}
\def\x2{{\xi }_{yy}}
\def\x3{{\xi }_{xy}}

\def\e1{{\eta }_{xx}}
\def\e2{{\eta }_{yy}}
\def\e3{{\eta }_{xy}}

\newcommand{\ds}{\displaystyle }
\newtheorem{theorem}{Theorem}
\newtheorem{lemma}{\bf Lemma}

\newcommand{\beqn}{\begin{eqnarray*}}
\newcommand{\eeqn}{\end{eqnarray*}}
\newcommand{\beqnn}{\begin{eqnarray}}
\newcommand{\eeqnn}{\end{eqnarray}}
\newcommand{\p}{\partial}
\newcommand{\bb}{\begin{equation}}
\newcommand{\ee}{\end{equation}}
\newcommand{\ba}{\begin{array}}
\newcommand{\ea}{\end{array}}
\newcommand{\R}{\mathbb{R}}

\begin{document}
\pagenumbering{arabic}
\title{\huge \bf Group Classification of Burgers' Equations}
\author{\rm \large Igor Leite Freire \\
\\
\it Instituto de Matem\'atica,
Estat\'\i stica e \\ \it Computa\c c\~ao Cient\'\i fica - IMECC \\
\it Universidade Estadual de Campinas - UNICAMP \\ \it C.P.
$6065$, $13083$-$970$ - Campinas - SP, Brasil
\\ \rm E-mail: igor@ime.unicamp.br}
\date{\ }
\maketitle
\vspace{1cm}
\begin{abstract}
In this work we carry out a complete group classification of Burgers'
equations.
\end{abstract}
\vskip 1cm
\begin{center}
{2000 AMS Mathematics Classification numbers:\vspace{0.2cm}\\
22E60, 35K55, 35Q53, 58J70\vspace{0.2cm} }
\end{center}
\pagenumbering{arabic}
\newpage

\section{Introduction}

Let $x\in M\subseteq\mathbb{R}^{n},\, M$ open, $u:M\rightarrow\mathbb{R}$ a smooth
function and $k\in\mathbb{N}$. We use $\partial^{k} u$ to denote the jet bundle corresponding to all $k$th partial derivatives of $u$ with respect to $x$. We simply denote $\p^{1}u$ by $\p u$. 

Partial differential equations are used to model many different kinds of phenomena in science and engeneering. Linear equations give mathematical description for physical, chemical or biological processes in a first approximation only. In order to have a more detailed and precise description a mathematical model needs to incoporate nonlinear terms. Nonlinear equations are difficult to solve analytically. However, in the end of century $XIX$ Sophus Lie developed a method that is widely useful to obtain solutions of a differential equation. This method is currently called \textit{Lie point symmetry theory}. Some applications of this method in (nonlinear) differential equations can be found in \cite{bl, yi,igi, ib, la,me, ol, ou}.

Lie used group properties of differential equations in order to actually solve them, i.e., to construct their exact solutions. Nowadays symmetry reductions are one of the most powerful tools for solving nonlinear PDEs.

A Lie point symmetry\footnote{In fact, a Lie point symmetry is given by the exponential map $(\exp{S})(x,u)=:(x^{\ast},u^{\ast})\in\R^{n}\times\R$. We are identifying the point transformation with its generator.} of a PDE $F=F(x,u,\p u, \cdots, \p^{m} u)=0$ of order
$m$ is a vector field
\bb\label{simetria}
S=\xi^{i}(x,u)\frac{\partial}{\partial x^{i}}+\eta(x,u)\frac{\partial
}{\partial u}%
\ee
on $M\times\mathbb{R}$ such that $S^{(m)}F=0$ when $F=0$ and
\[
S^{(m)}:=S+\eta^{(1)}_{i}(x,u,\partial u)\frac{\partial}{\partial u_{i}}%
+\cdots+\eta^{(m)}_{i_{1}\cdots i_{m}}(x,u,\partial u,\cdots,\partial^{m}
u)\frac{\partial}{\partial u_{i_{1}\cdots i_{m}}}%
\]
is the extended symmetry on the jet space $(x,u,\partial u,\cdots,
\partial^{k} u)$.

The functions $\eta^{(j)}(x,u,\partial u,\cdots,\partial^{j} u)$, $1\leq j\leq
m$, are given by
\bb\label{etas}
\begin{array}
[c]{l c l}%
\eta^{(1)}_{ i} &: = & D_{i}\eta-(D_{i}\xi^{j})u_{j},\\
&  & \\
\eta^{(j)}_{i_{1}\cdots i_{j}} &: = & D_{i_{j}}\eta^{(j-1)}_{i_{1}\cdots
i_{j-1}}-(D_{i_{j}}\xi^{l})u_{i_{1}\cdots i_{j-1}l},\;2\leq j\leq m,
\end{array}
\ee
where
$$D_{i}:=\f{\p}{\p x^{i}}+u_{i}\f{\p}{\p u}+u_{ij}\f{\p}{\p u_{j}}+\cdots+u_{ii_{1}\cdots i_{m}}\f{\p}{\p u_{i_{1}\cdots i_{m}}}+\cdots $$
is the \textit{total derivative operator}. We shall not present more preliminaries concerning the Lie point symmetries of
differential equations supposing that the reader is familiar with the basic notions and methods of group analysis \cite{bl,ib,ol}.

In a previous paper, Lagno and Samoilenko \cite{la} made the group classification of quaselinear evolution equation
\bb\label{hqgen}
u_{t}=F(x,t,u,u_{x}, u_{t})u_{xx}+G(x,t,u,u_{x},u_{t}),
\ee
where $u=u(x,t)$, for general smooth functions $F$ and $G$.

When $F=1\text{ and }G=-u\,u_{x}$, the equation is communly known like Burgers' equation because it was first studied by Burgers in the last century. 

In this article we shall call Burgers' equation the general case of equation (\ref{hqgen}) when $F=\nu=const$ and $G=-g(u)u_{x}$, where $g(u)$ is a smooth function. This is the same terminology used in \cite{me}. 

In \cite{ou}, the group classification of (\ref{hqgen}) is carried out with $F=0$ and \text{$G=-g(u)u_{x}$}, for particular choices of the function $g$. In \cite{me}, the results obained in \cite{ou} are generalized for arbitrary $g(u)$. These equations are called \textit{inviscid Burguers' equations}.

In this paper we are interested in generalize the group classification obtained in \cite{ou, me} to the Burgers' equation with $\nu> 0$ and $g(u)$ arbitrary. The linear case $g(u)=k=const$ shall not be considered here because we are interested only in nonlinear cases. To the particular case $g(u)=0$, see the group classification in \cite{bl,ol, igi}. To the Burgers' equation $u_{xx}=u_{t}+uu_{x}$, the Lie point symmetries can be found in \cite{bl, ol}. However, we shall present this case in this article for completeness.

The remaining of the paper is organized as follows. In section 2 we carry out the complete group classification of equation
$$
\nu u_{xx}=u_{t}+g(u)u_{x},
$$
and in the section 3 we identify the classical Lie algebras that the symmetry Lie algebras are isomorphic.

\section{Main result}

\

Let us consider the equation

\bb\label{bu}
\nu u_{xx}=u_{t}+g(u)u_{x},
\ee
with $\nu> 0$ and $g'(u)\neq 0$. In the remaining of this paper, we shall be supposing that all functions are smooths and they are well defined.

\begin{lemma}\label{lem1}
Let 
\bb\label{sym}
S=\xi(x,t,u)\f{\p}{\p x}+\phi(x,t,u)\f{\p}{\p t}+\eta(x,t,u)\f{\p}{\p u}
\ee
be a symmetry of equation $(\ref{bu})$. Then $\xi=\xi(x,t)$, $\phi=\phi(t)$ and $\eta=\al(x,t)u+\be(x,t)$.
\end{lemma}

\begin{proof}
From \cite{b1,bl, igi}, we conclude that $\xi=\xi(x,t)$, $\phi=\phi(x,t)$ and $\eta=\al(x,t)u+\be(x,t)$. From \cite{la}, $\phi=\phi(t)$.
\end{proof}

\begin{lemma}\label{deteq}
The linearly independet set of determing equations of equation $(\ref{bu})$ is:
\bb\label{eq1}
\phi'(t)=2\xi_{x}
\ee

\bb\label{eq2}
u\al_{t}-\nu u\al_{xx}+ug(u)\al_{x}+\be_{t}-\nu\be_{xx}+g(u)\be_{x}=0,
\ee

\bb\label{eq3}
\xi_{t}+2\nu\al_{x}-g(u)\xi_{x}-u g'(u)\al-g'(u)\be=0.
\ee
\end{lemma}

\begin{proof}
It follows from the invariance condition $S^{(2)}F=0$ whenever $$F= \nu u_{xx}-u_{t}-g(u)u_{x}=0.$$ 
See also \cite{la, grego1, grego2}.
\end{proof}

\begin{theorem}\label{main}\texttt{Group Classification Theorem}\\
\\
The widest Lie point symmetry group of Burgers' equation $(\ref{bu})$ with an arbitrary $g(u)$, is
determined by the operators
\bb\label{sg}
X=\f{\p}{\p x},\;\;\;T=\f{\p}{\p t}.
\ee
For some special choices of the function $g(u)$ it can be extended in the
cases listed below. We shall write only the generators additional to $(\ref{sg})$.

\begin{enumerate}
\item If $g(u)=u$, then
$$
\ba{c}
\ds{B_{11}=tx\f{\p}{\p x}+t^{2}\f{\p}{\p t}+(x-tu)\f{\p}{\p u},\;\;\;B_{12}= t\f{\p}{\p x}+\f{\p}{\p u}},\\
\\
\ds{B_{13}=x\f{\p}{\p x}+2t\f{\p}{\p t}-u\f{\p}{\p u}}.
\ea
$$
\item If $\ds{g(u)=u^{p},\;p\neq 0,\;1}$, then the additional generator is
$$B_{2}=x\f{\p}{\p x}+2t\f{\p}{\p t}-\f{1}{p}u\f{\p}{\p u}.$$

\item If $\ds{g(u)=\log{u}}$, then the additional generator is
$$B_{3}=t\f{\p}{\p x}+u\f{\p}{\p u}.$$

\item If $\ds{g(u)=e^{bu}},\,b=const\neq 0$, then
$$B_{4}=x\f{\p}{\p x}+2t\f{\p}{\p t}-\f{1}{b}\f{\p}{\p u}.$$

\item If $\ds{g(u)=\f{1-u}{1+u}}$, then
$$B_{5}=(x- t)\f{\p}{\p x}+2t\f{\p}{\p t}+(1+u)\f{\p}{\p u}.$$

\item If $\ds{g(u)=\f{ 1}{1+u}}$, then
$$B_{6}=x\f{\p}{\p x}+2t\f{\p}{\p t}+(1+u)\f{\p}{\p u}.$$

\item If $\ds{g(u)=\f{ u}{1+u}}$ or $\ds{g(u)=\f{u}{1-u}}$, then the additional generator is
$$B_{7}=(x+t)\f{\p}{\p x}+2t\f{\p}{\p t}+(1+u)\f{\p}{\p u}.$$

\end{enumerate}
\end{theorem}

\begin{proof}
If $g$ is an arbitrary function, in order to equations (\ref{eq2}) and (\ref{eq3}) be true, necessarily we have $\xi_{t}-2\nu\al_{x}=\xi_{x}=\al=\be=0$. Then, from equations (\ref{eq1}) and (\ref{eq3}) we conclude that $\xi=c_{1}=const$ and $\phi=c_{2}=const$. Then, the symmetry (\ref{sym}) is spanned by translations in $x$ and $t$.

The proof of case $g(u)=u$ can be found in \cite{bl, ol}. To other cases, substituting the functions listed in the Theorem in equations (\ref{eq2}) and (\ref{eq3}), we obtain an identity in terms of $u,\, g(u),\, g'(u)$ and $ug'(u)$. Solving it, we obtain the coefficients $\xi, \phi,\al$ and $\be$ of symmetry (\ref{sym}).
\end{proof}

\section{Symmetry Lie algebras}

\

In this section we are interested in classify the symmetry Lie algebras of equation (\ref{bu}). In the next theorem, we present only the non-null Lie brackets.

\begin{theorem}\label{liealg}
The symmetry Lie algebras of the Burgers' equation are
\begin{enumerate}
\item If $g(u)= u$, then
$$
[X,B_{11}]=B_{12},\;\; [X,B_{13}]=X,\;\;  [T,B_{11}]=B_{13}, 
$$
$$
[T,B_{12}]=X,\;\; [X,B_{13}]=2T,\;\; [B_{11},B_{13}]=-2B_{11},\;\; [B_{12},B_{13}]=-B_{12}.
$$

\item If $\ds{g(u)=u^{p},\;p\neq 0,\;1}$, then 
$$[X,B_{2}]=X,\;\;[T,B_{2}]=2T.$$

\item If $\ds{g(u)=\log{u}}$, then 
$$[T,B_{3}]=X.$$

\item If $\ds{g(u)=e^{bu}},\,b=const$, then
$$[X,B_{4}]=X,\;\;[T,B_{4}]=2T.$$

\item If $\ds{g(u)=\f{1-u}{1+u}}$, then
$$[X,B_{5}]=X,\;\;[T,B_{5}]=-X+2T.$$

\item If $\ds{g(u)=\f{ 1}{1+u}}$, then
$$[X,B_{6}]=X,\;\;[T,B_{6}]=2T.$$

\item If $\ds{g(u)=\f{ u}{1\pm u}}$, then
$$[X,B_{7}]=X,\;\;[T,B_{7}]=X+2T.$$

\end{enumerate}

\end{theorem}

Let $\text{\goth{g}}_{1}:=\{X,T,B_{11},B_{12},B_{13}\}$ and $\text{\goth{g}}_{i}:=\{X,T,B_{i}\}\,2\leq i\leq 7$.

It is immediate that $\text{\goth{g}}_{2}\cong \text{\goth{g}}_{4}\cong \text{\goth{g}}_{6}$ and, under the change $X\mapsto -X$, $\text{\goth{g}}_{5}\cong \text{\goth{g}}_{7}$. 
\begin{theorem}\label{iso}
$\text{\goth{g}}_{2}\cong \text{\goth{g}}_{5}.$
\end{theorem}
\begin{proof}
Let $e_{1}:=X,\,e_{2}:=X+T$ and $e_{3}:=B_{2}$. Then, \bb\label{a12}[e_{1},e_{3}]=e_{1} \text{ and } [e_{2},e_{3}]=2e_{2}.\ee
\end{proof}

Under the change $e_{1}'= e_{2},\,e_{2}'= e_{1}$ and $e_{3}'=\f{1}{2}e_{3}$ in (\ref{a12}), the following result is a consequence from Theorems \ref{liealg}, \ref{iso} and \cite{pswz,pw}.

\begin{theorem}
$\text{\goth{g}}_{1}\cong A_{5,40}$, $\text{\goth{g}}_{2}\cong \text{\goth{g}}_{4}\cong \text{\goth{g}}_{5}\cong \text{\goth{g}}_{6}\cong \text{\goth{g}}_{7}\cong A_{3,5}^{\f{1}{2}}$, $\text{\goth{g}}_{3}\cong A_{3,1}$, where $A_{3,1}$ is the Weyl-Heisenberg algebra (see \text{\cite{yi}}).
\end{theorem}

\section{Acknowledgments}
The author would like to thank Lab. EPIFISMA (Proj. FAPESP) for
 having given him the opportunity to use excellent computer facilities.

\end{document}